\documentclass{aastex}
\usepackage{graphicx,emulateapj5,natbib}
\def\myputfigure#1#2#3#4#5%
{\vskip#5pt\makebox[0pt]{\hskip#2in
\includegraphics[width=#3\textwidth]{#1}}\vskip#4pt\hfill}

\def\beb{\begin{thebibliography}{999}}
\def\eeb{\end{thebibliography}}

\def\be{\begin{equation}}
\def\ee{\end{equation}}
\def\bea{\begin{eqnarray}}
\def\eea{\end{eqnarray}}
\def\o{\over}

\def\om{$\Omega_M~$}
\def\gamx{$\gamma_X$}
\def\gams{$\gamma_{SZ}$}
\def\pcl{$P_{cl}(k)$}

\begin{document}

\shorttitle{Self Calibration in Cluster Studies of Dark Energy}
\shortauthors{Majumdar and Mohr}
\submitted{submitted to ApJ May 16, 2003}

\title{Self Calibration in Cluster Studies of Dark Energy:  Combining the Cluster Redshift Distribution, the Power Spectrum and Mass Measurements}

\author{Subhabrata Majumdar\altaffilmark{1}  and Joseph J Mohr\altaffilmark{1,2}}
\altaffiltext{1}{Department of Astronomy, University Of Illinois, 1002 West Green St., Urbana, IL 61801}
\altaffiltext{2}{Department of Physics, University Of Illinois, 1002 West Green St., Urbana, IL 61801}
\email{subha@uiuc.edu}\email{jmohr@uiuc.edu}

\begin{abstract}
We examine the prospects for measuring the dark energy equation of state parameter $w$ within the context of any uncertain redshift evolution of galaxy cluster structure \citep[building on][]{majumdar03} and show that including the redshift averaged cluster power spectrum, ($\bar{P}_{cl}$), and direct mass measurements of 100 clusters helps tremendously in reducing cosmological parameter uncertainties. Specifically,  we show that when combining the redshift distribution and the power spectrum information for a particular X--ray survey (DUET) and  two SZE surveys (SPT \&  $Planck$), the constraints on the dark energy equation of state $w$ can be improved by roughly a factor of 4.  Because surveys designed to study the redshift distribution of clusters will have all the information necessary to construct  $\bar{P}_{cl}$, the benefit of adding $\bar{P}_{cl}$ in reducing uncertainties comes at no additional observational cost.  Combining  detailed mass studies of $100$ clusters with the redshift distribution improves the parameter uncertainties by a factor of 3-5.  The data required for these detailed mass measurements-- assumed to have 1$\sigma$ uncertainties of 30\%-- are accumulating in the the {\it XMM-Newton} and {\it Chandra} archives.  The best constraints are obtained when one combines both the power spectrum constraints and mass measurements with the cluster redshift distribution;  when using the survey to extract the parameters and evolution of the mass--observable relations, we estimate the uncertainties on $w$ of $\sim$4\% to 6\%.  These parameter constraints are obtained from self-calibrating cluster surveys alone.  In combination with CMB or distance measurements that have different parameter degeneracies, cluster studies of dark energy will provide enhanced constraints and allow for cross--checks of systematics.
\end{abstract}


\section{Introduction}
Galaxy clusters have been used extensively to determine the cosmological matter density parameter and the amplitude of density fluctuations.  
The number density of massive clusters is exponentially sensitive to the amplitude of initial Gaussian fluctuations [whose normalization we describe using
$\sigma_8$, the {\it rms} fluctuations of overdensity within spheres of 8$h^{-1}$~Mpc radius]  \citep{eke96, henry97, bahcall98, viana99, reiprich02, lin03}.   In principle, clusters contain information about many aspects of the evolution of the universe.
 \citet{wang98} argued that a measurement of the changes of cluster number density or abundance with redshift would provide constraints on the dark energy equation of state parameter $w\equiv p/\rho$.  \citet{haiman01} showed that with future large surveys it should be possible to obtain precise measurements of the amount $\Omega_E$ and nature $w$ of the dark energy.   Since then, a series of papers has focused on the prospects of high yield cluster
surveys as probes of dark energy \citep{holder01b, weller01, levine02, hu03a, majumdar03,hu03b}. These
 papers also show that constraints from cluster surveys will be complimentary to those from cosmic microwave background (CMB) anisotropy and SNe Ia distance measurements.

The potential of using cluster surveys can only be realized fully if we not only know the cluster redshifts but also have some estimate of the cluster masses \citep{haiman01}. However, this total mass has to be inferred indirectly through use of cluster scaling relations and observables like the X-ray flux, SZE flux, galaxy light and weak lensing shear.   Recently, \citet{levine02} examined an X--ray cluster survey, showing that a sufficiently large survey allows one to measure cosmological parameters and constrain the cluster mass--observable relation simultaneously if one assumes perfect knowledge of the redshift evolution of galaxy cluster structure. If one allows for an imperfect understanding of galaxy cluster evolution, then the constraints on $w$ are seriously weakened \citep[hereafter MM03]{majumdar03}. 
To overcome this important limitation, MM03 examined the impact of follow-up of a fraction of the survey cluster sample and 
showed that even a $1\%$ follow-up helps in improving the $w$ constraints by a factor 2 to 3.  
These calculations underscore the importance of incorporating information from multiple observables into future cluster surveys, and they demonstrate that cluster surveys are essentially self--calibrating-- containing enough  information to solve for the mass--observable relation at every redshift and constrain cosmological parameters.   As another solution to overcome the evolution-cosmology degeneracy, \citet {hu03b} has shown that there is potential for internal calibration by using the shape of the mass function of clusters in redshift slices.   In the case of tight mass--observable relations and accurate redshifts, this approach essentially enables one to solve for the mass observable relation independently in each redshift slice of the survey.  Additional calculations are needed to examine how well this very promising approach  will work in future cluster surveys.

Here we look at another way to improve constraints on the equation of state of dark energy. We show how one may 
use the redshift averaged cluster power spectrum $\bar{P}_{cl}(k)$, derived from the correlated positions of galaxy clusters in the survey,
to tighten the constraints on cosmological parameters. Because the sky position and the redshift 
(already needed to measure the redshift distribution) for each cluster gives us enough information to construct $\bar{P}_{cl}$, one essentially gets the cluster power spectrum {\it without further observational effort}. 
The amplitude and shape of the cluster power spectrum depends on the underlying dark matter power
spectrum and the bias of the tracers, which is related to the mass of the halo \citep{mo96,sheth99} in the same way as the mass function of the halos. 
Thus, a measure of biased cluster power spectrum essentially provides an indirect measurement of the mass  of the cluster tracers . This additional information is essential in constraining the evolution of the mass-observable relation, which is key to obtaining tight cosmological constraints from the cluster redshift distribution.  In addition, the shape of the cluster power spectrum provides an additional constraint on the matter density because of its effect on the transfer function.

Moreover,  an advantage of $\bar{P}_{cl}$ over any spatial correlation, $\xi(r)$, studies is in the respective uncertainty analysis. For $\bar{P}_{cl}$, one can assume, for an
initial Gaussian field, the statistical fluctuation to be mainly uncorrelated \citep{fkp}. However, $\bar{P}_{cl}$ is similar to $\xi(r)$
in the sense that both exhibit different parameter degeneracies than those from using $dN/dz$ and hence combining power spectrum constraints with the cluster redshift distribution helps in tightening parameter constraints.  
The aim of this paper is to examine the complementarity of the redshift distribution and cluster power spectrum that can be obtained in any high yield cluster surveys.  We show that by combining these two analyses,  dramatic improvement on $w$-constraints can be realized even in the face of uncertain evolution of the cluster mass--observable relation.

The paper is arranged in the following way. In $\S$\ref{sec:prelim} we describe the basic formalism and then in $\S$\ref{sec:fisher} we describe three representative surveys and our method of estimating 
parameter constraints. In $\S$\ref{sec:results} we present our results and finally conclusions are drawn in
$\S$\ref{sec:conclusions}.

\section{Preliminaries}
\label{sec:prelim}
\subsection{Cluster Redshift Distribution}
To begin, we estimate the redshift distribution of detectable clusters within a survey solid angle $\Delta \Omega_M$,
\be
{{dN}\o{dz}} ~ = ~ \Delta\Omega {{dV}\o{dzd\Omega}}(z)\int_{0}^\infty f(M,z) {{dn(M,z)}\o{dM}}dM
\label{eqn:dndz}
\ee
where ${{dV}/{dzd\Omega}}$ is the comoving volume element, $({{dn}/{dM}})dM$ is the comoving density of clusters of mass $M$, and $f(M,z)$ is the redshift dependent cluster selection function for the survey.  In this analysis we take $f(M,z)$ to be a step function at some limiting mass $M_{lim}$, which corresponds to the mass of a cluster that lies at the survey detection threshold.   In a real survey the selection function will depart from a step function, and this has been examined elsewhere \citep{holder00,lin03}. We use the cluster mass function ${{dn}/{dM}}$ determined from structure formation simulations \citep{jenkins01}.  Sample $dn/dz$ plots can be found in MM03.

For an X-Ray survey, the mass-observable relation takes the form of bolometric flux - mass relations
and we use 
\be
f_x(z)4\pi d_L^2 = A_x M_{200}^{\beta_x} E^2(z)\left(1+z\right)^{\gamma_x}
\label{eq:lx-m}
\ee
where $f_x$ is the observed flux in units of erg~s$^{-1}$cm$^{-2}$, $d_L$ is in units of Mpc, $M_{200}$  in units of $10^{15}M_\odot$ is the mass enclosed within a radius $r_{200}$ having a overdensity of $200$  with respect to critical and  $H(z)=H_0E(z)$. 
where $E^2(z)=\Omega_M(1+z)^3+\Omega_k(1+z)^2+\Omega_E^{3(1+w)}$.    
We  convert $M_{200}$ to $M(z)$, the halo mass appropriate for our mass function at redshift $z$ using a halo model \citep[][hereafter, NFW]{navarro97}. Following MM03, the possibility of non-standard evolution of the
 mass--observable relation is included through the parameter $\gamma_x$.   
In our fiducial model we take $\gamma_x=0$ to be consistent with the observed weak evolution in the 
luminosity--temperature relation \citep{vikhlinin02}, and we choose $\beta_x=1.807$ and 
$\log{A_x}=-3.926$ \citep{reiprich02}.

For an SZE survey, the corresponding SZE flux-mass relation is given by
\be
f_{sz}(z,\nu) d_A^2 ={f(\nu)f_{ICM}}A_{sz}{{M^{\beta_{sz}}_{200}}} E^{3/2}(z)\left(1+z\right)^{\gamma_{sz}}
\label{eq:fsz-m}
\ee
where $f(\nu)$ is the frequency dependence of the SZE distortion, $f_{sz}$ is the observed flux in mJy, $M_{200}$ is in units of $10^{15}M_\odot$, $f_{ICM}=0.12$ \citep[e.g][]{mohr99} and $d_A$ is in units of Mpc  \citep[see][]{diego01}.    For our fiducial
model, we use the mass - temperature relation for clusters having $T > 3$KeV \citep{finoguenov01}, to get $\log{A_{sz}}=16.47$, $\beta_{sz}=1.68$ and $\gamma_{sz}=0$ to model standard structure evolution. 
Note that in determining the estimated uncertainties on cosmological parameters, we allow the 
normalization of these mass--observable relations to be free to vary.  Finally, we impose a
minimum cluster mass of $10^{14}h^{-1}$~M$_\odot$ on all surveys. 

\subsection{The Cluster Power Spectrum}
The power spectrum of future high yield surveys provides a different probe of the background
cosmological model. Taken alone, the constraints from cluster power spectrum are far weaker than those from $dN/dz$ alone. 
However, when
combined with other measures of cosmological parameters, the power spectrum studies
can give very powerful constraints. While in most cases, people have studied the power spectrum
for galaxy surveys, the cluster power spectrum holds similar promise because clusters
are highly biased when compared to galaxies, making it possible to obtain similar statistical uncertainties with far smaller samples. Another important issue that contrasts the cluster power
spectrum, \pcl,  from that of galaxies is the fact that the bias, $b(M,z)$,  for clusters can be
determined from large scale N-body simulations and theoretical calculations \citep{mo96,sheth99}.  In addition, cluster masses are related to simple observables, making it possible to directly connect the bias of the cluster power spectrum to these same observables.  The same physics of gravitational collapse of dark matter dominated halos underlies the theoretical and numerical predictions of the bias--mass relation and the mass function of collapsed haloes \citep{mo96,sheth99,jenkins01,sheth01}.

\myputfigure{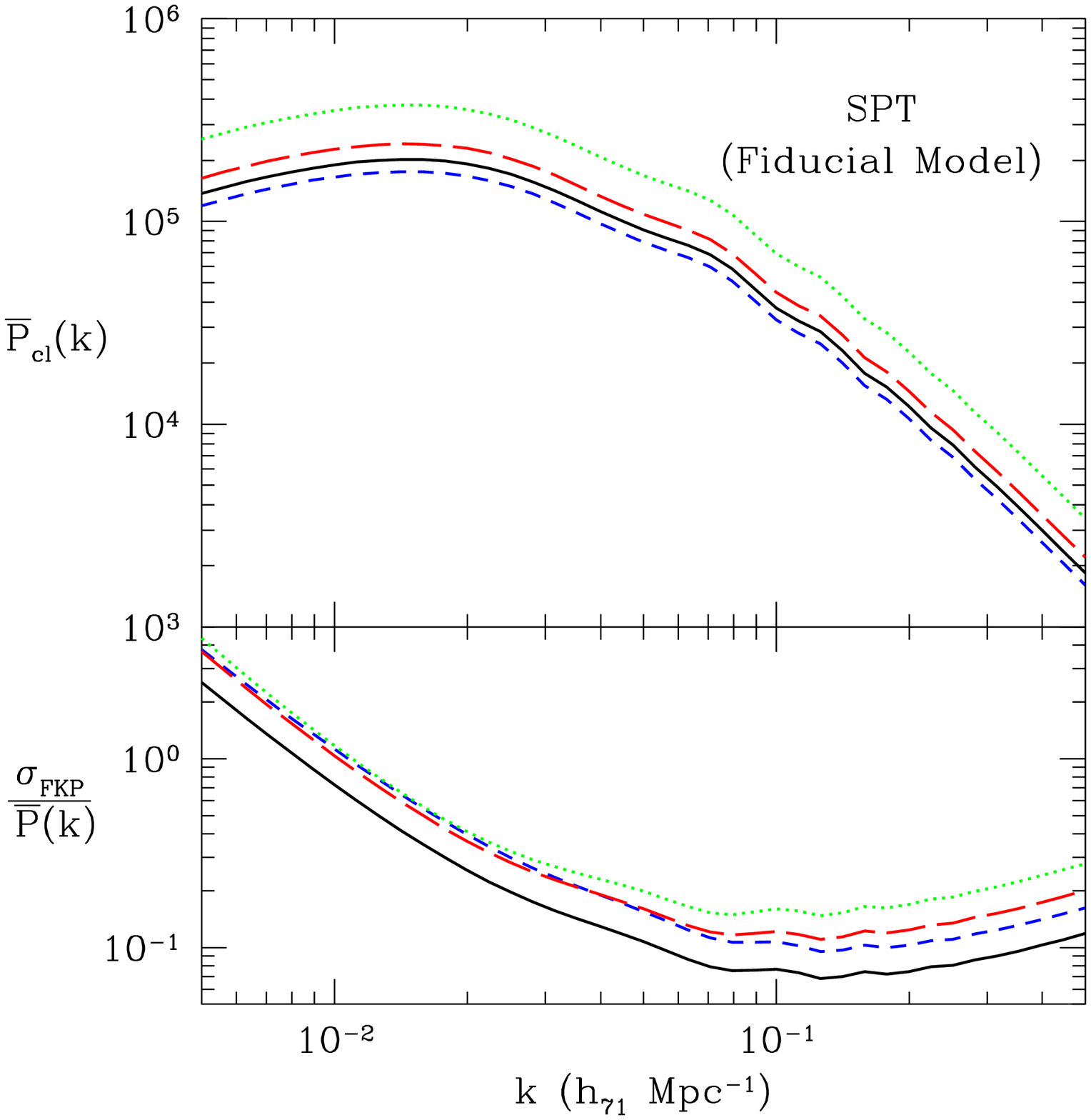}{3.0}{0.50}{-15}{-05}
\figcaption{The upper panel shows redshift averaged cluster power spectrum, $\bar{P}_{cl}$, for the SPT survey, for the three
redshift bins, $z=0-0.42$ (short-dashed line), $z=0.42-0.72$ (long-dashed) and $z=0.72-3$ (dotted line), each bin having 
$\sim 9700$
clusters. The solid line shows
the results for one single redshift bin from $z=0-3$. The lower panel shows the fractional uncertainties in estimating the power spectrum,
$\sigma_{\bar{P}_{cl}}(k)/\bar{P}_{cl}(k)$, using the method detailed in \citet{fkp}.
\label{fig:pk}\vskip5pt}

To calculate the parameter sensitivity of \pcl we follow the method described in a number of papers dealing with galaxy power spectrum \citep{tegmark97a, eisenstein99}, keeping in mind
that clusters are biased with respect to the underlying matter distribution.  We use the bias model in
\citet{sheth99}.  
 In the limit where the survey volume is much larger than the scale of any features in the power
spectrum \citep{hamilton97}, the uncertainties can be well estimated with the method given by
 \citet[hereafter FKP]{fkp}. The cluster power spectrum is constructed  as 
$P_{cl}(k,z) = b^2_{eff}(z)P(k,z)$ where $P(k,z)$ is the matter power spectrum and $b_{eff}$ at a 
particular redshift is the cluster mass function averaged linear bias, $b(M,z)$ \citep{suto00,Moscardini00}. 
The effective bias is, then, given by
\be
b_{eff}(z) = {{\int dM {{dn}\over{dM}}b(M,z)}\over{\int dM {{dn}\over{dM}} }}
\ee
where $dn/dM$ is the cluster mass function obtained from N-body simulations \citep{sheth99, jenkins01}.
This is
 valid for a small redshift interval $\delta z$ at a particular $z$.  It is useful
to measure the power spectrum over a wide redshift range. In this case, the light cone averaged
cluster power spectrum, $\bar{P}_{cl}$, in the redshift range of interest between $z_{min}$ and $z_{max}$
is obtained as
\be 
\bar{P}_{cl}(k) = {{ \int^{z_{max}}_{z_{min}}  dz {dV\o dz} n^2(z) P_{cl}(k,z) } \o
    {\int^{z_{max}}_{z_{min}}  dz {dV\o dz} n^2(z) }}
\ee
where the comoving number density of clusters in the flux limited surveys is computed by 
integrating Eqn \ref{eqn:dndz}.
Thus, $\bar{P}_{cl}$ not only depends on cosmology but also
on the specific survey that determines the redshift dependence of observed clusters. 

We assess the sensitivity of $\bar{P}_{cl}$ to different parameters through the Fisher matrix formalism detailed in the next section.  
Note  that the evolution of $b(M,z)$ follows the growth of structures in hierarchical models of structure formation  
\citep{mo96, sheth99,sheth01}. Also note that the amplitude of the power spectrum
evaluated on the light cone for a given flux limit increases with $z$; the increasing bias at higher redshift offsets the reduction in amplitude of the underlying power spectrum of dark matter fluctuations.  For all the surveys, the power
spectrum is calculated in the $k$-range 0.0014$h^{-1}$Mpc$^{-1}$ to 1.4$h^{-1}$Mpc$^{-1}$. The uncertainty in the estimation of $\bar{P}_{cl}$ increases rapidly as we decrease $k$, and 
effectively cuts off the contribution to the Fisher matrix for wavenumbers less than $k_{min}$. At higher
$k$, non-linear effects become important and increase $\bar{P}_{cl}$.  Although, the power spectrum in the non-linear
regime remains a well defined quantity, it is unwise to to use the Fisher matrix approach (see below) to estimate constraints on
parameters as the density field becomes nonlinear and non--Gaussian; in this regime Equation \ref{eq:fishpk} could give incorrect estimates of the parameter constraints. One needs to use Monte-Carlo methods to accurately study processes having non-Gaussianities.
We ask the reader to keep these considerations in mind when assessing our results. 
Finally, we neglect redshift space distortion and non-linear contributions to $\bar{P}_{cl}$,  which produce a small change in the power spectrum amplitude \citep{suto00}.

\subsection{Follow-up of Large Solid Angle Surveys}
Finally, MM03 introduced the idea of follow-up of a fraction of the cluster sample to calibrate
the mass-observable relation and highlighted the resulting improvement in parameter
constraints. Followup requires detailed X--ray, SZE or galaxy spectroscopic observations that allow one to measure the mass-like quantity $M_f(\theta)=M(\theta)/d_A$, which we will refer to as the follow-up mass, where the $M(\theta)$
can be calculated using a NFW profile for the matter distribution. We assume a concentration $c=5$.   
At fixed redshift, this follow-up  produces an $M_f$--$f_x$ of $M_f$--$f_{sz}$ relation which would provide direct constraints on the structural evolution of the clusters.
The parameter sensitivity of these scaling relation observations can exhibit quite different degeneracies than for the cluster redshift distribution, making the scaling relations and the cluster redshift distribution complementary (see MM03).  We emphasize that a large number of clusters are now being studied in great detail with Chandra, XMM and deep optical weak lensing data.  This growing ensemble of well studied clusters will be available for mass calibration in all future surveys.  We model this additional information by placing 100 clusters uniformly over a range of redshifts from $0.3\le z\le 1.2$ and masses from 
$10^{14}h^{-1} M_\odot$ to $10^{15}h^{-1} M_\odot$.

\section{Cosmological Sensitivity of a Survey}
\label{sec:fisher}
\subsection{Fisher Matrix Technique}
  We employ the Fisher matrix technique to probe the relative sensitivities of  cluster surveys to different cosmological and cluster structural parameters. The Fisher matrix information for a data set \citep[see][]{tegmark97b,eisenstein98b} is defined as 
$F_{ij} \equiv -<{{\partial^2 \ln{\mathcal{L}}}\o{\partial p_i \partial p_j}}>$, where $\mathcal{L}$ is the likelihood for an observable (proportional to ${{dN}\o{dz}}$ for the redshift distribution, $\bar{P}_{cl}$ for the power spectrum and  $M_f $ for the follow-up) and $p_i$ describes our parameter set. The inverse $F_{ij}^{-1}$ describes the best attainable covariance matrix $[C_{ij}]$ for measurement of the parameters considered. The diagonal terms in $[C_{ij}]$ then give the uncertainties on each of our parameters.  
To estimate the full potential for cluster surveys, we calculate the final Fisher matrix is the sum of the Fisher matrices for the cluster
redshift distribution ($F_{ij}^{s}$), the cluster power spectrum ($F_{ij}^{p}$),  the
follow up observations ($F_{ij}^{f}$) and the priors.

We construct the survey Fisher matrix $F_{ij}^{s}$ following \citet{holder01b} as
\be
F_{ij}^{s} = \Sigma_n {{\partial N_n}\o{\partial p_i}} {{\partial N_n}\o{\partial p_j}}{{1}\o{{N_n}}},
\ee
where we sum over $n$ redshift bins of size $\Delta z = 0.01$ to $z_{max}=3.0$ and $N_n$ represents the number of surveyed clusters in each redshift bin $n$. 
The Fisher matrix for the power spectrum is given by 
\be
F_{ij}^{p} = \Sigma_m {{\partial \bar{P}_{cl}(k)}\o{\partial p_i}} {{\partial \bar{P}_{cl}(k)}\o{\partial p_j}} {{\Delta k_m}\o{{\sigma^2}_{FKP}}}
\label{eq:fishpk}
\ee
where we sum over $m$ logarithmic bins in $k$-space between $k_{min}$ and $k_{max}$ (see, for example \citet{tegmark97a}).
 The 
variance in each $k$-bin is calculated using the formulation given in \citet{fkp} with optimal 
weighting for each survey.

The Fisher matrix for the follow-up is constructed  as
\be
F_{ij}^{f} = \Sigma_n \left({{\partial M_f}\o{\partial p_i}}{{\partial M_f}\o{\partial p_j}}  {{1 }\o{{\sigma^2_{M_f}}}}\right)
\label{eq:fishfollow}
\ee
where $M_f$ is a function of halo mass $M$ and angular radius $\theta$ at a particular redshift $z$.  As noted above, we model the available followup information from deep X-ray studies of clusters by summing over 100 clusters that uniformly span a range of redshifts ($0.3\le z\le 1.2$) and a range of masses ($10^{14}h^{-1}M_\odot$ to $10^{15}h^{-1}M_\odot$).  In this way we attempt to model the information provided by  all the detailed cluster analyses that accumulate in the literature.  Any followup of clusters in a particular survey would result in additional constraints on cluster structural evolution.   
 In this analysis we take $\sigma_{M_f} = 0.3M_f$ to be the characteristic uncertainty in the follow-up mass measurements.  For more details on cluster follow-up specifications, please refer to MM03.

\subsection{Future Surveys, Fiducial Cosmology and External Priors}
\label{sec:fiducial}
For our study, we adopt three representative surveys, two in SZE and one in X-Ray. For the X-Ray
survey we examine the  $10^4$~deg$^2$ flux limited X--ray survey proposed as part of the DUET mission to the NASA Medium--class Explorer Program. For the SZE surveys, 
we consider the   4,000~deg$^2$ SZE survey to be carried out with an 8m telescope being constructed for the South Pole (SPT) and  the $Planck$ full sky SZE cluster survey. 
We model the DUET X-ray survey as having a bolometric flux limit of $f_x>1.25\times10^{-13}$~erg/s/cm$^2$ (corresponding to $f_x>5\times10^{-14}$~erg/s/cm$^2$ in the 0.5:2~keV band).  For our fiducial cosmological model,this survey yields $\sim$22,000 detected clusters, consistent with the known X-ray $\log N$--$\log S$ relation for clusters \citep[e.g.][]{gioia01}.  
We model the SPT SZE survey as a flux limited survey with $f_{SZ}>5$~mJy at 150~GHz.  Within our fiducial cosmological model this survey would yield $\sim 29,000$ clusters with measured fluxes. And finally for the $Planck$ survey, we take a flux limit of $f_{SZ}>50$~mJy at 353~GHz (at which the angular resolution of $Planck$ would reach 5 arc min) 
\citep{diego02}  which for 70\% of the sky yields $\sim$21,000 clusters.

The fiducial cosmological parameters (and priors) of our model are adopted from the first year WMAP
result \citep{bennet03} and are $h=0.71$, $\Omega_M=0.27$, $n=0.93$, $\Omega_B=0.044$, $\sigma_8= 0.84$ and $w=-1$.
From WMAP results we know that the curvature is very close to zero, 
and hence we assume a flat universe, taking $\Omega_{tot}=\Omega_M+\Omega_{E}=1$. 
We also neglect any variation in $w$ with redshift. 
The priors used for our calculations are $\sigma_h=0.035, \sigma_{\Omega_b}=0.004$ 
and $\sigma_n=0.03$.
 We do not use any priors on \om, $\sigma_8$ or $w$ for the 
main results of this paper as summarized in Table 1. However, to show
the effect of strong CMB priors on $\Omega_M, \sigma_8$ and $w$ 
(shown within parentheses in Table 1), we use
$\sigma_{\Omega_M}=0.04, \sigma_{\sigma_8}=0.04$ and $\sigma_w=0.11$.  Note that this provides a crude estimate of the complementarity with CMB constraints, because we have not used the parameter covariance appropriate for CMB power spectrum analyses. 

For power spectrum calculations, each survey is divided into three redshift bins such that each bin
has an approximately equal number of clusters. For example, for the SPT survey which has the most
clusters at high redshift, the  redshift binnings are $z=0.0-0.42, 0.42-0.72$  and $0.72-3.0$. 
For the fiducial cosmology and SPT survey, $\bar{P}_{cl}$ and the
the fractional uncertainty, $\sigma_{FKP}/{\bar{P}(k)}$,  for the three redshift bins are shown in 
Figure \ref{fig:pk} using short-dashed, long-dashed and dotted lines respectively. The solid line is for the
case of a single redshift bin from $z=0.0-3.0$.

\begin{table*}[htb]\small
\caption{Estimated Parameter Constraints\label{tab:params}}
\hbox to \hsize{\hfil\begin{tabular}{ccccccc}
Survey/Parameter		& \multicolumn{1}{c}{Only Cosmology} & \multicolumn{1}{c}{Std Evol.} & \multicolumn{1}{c}{Self-Cal} & \multicolumn{1}{c}{Self-Cal} & \multicolumn{1}{c}{Self-Cal} & \multicolumn{1}{c}{Self-Cal}  \\
		& \multicolumn{1}{c}{$dn/dz$} & \multicolumn{1}{c}{$dn/dz$}&\multicolumn{1}{c}{$dn/dz$} & \multicolumn{1}{c}{$dn/dz$+ $\bar{P}_{cl}$} &\multicolumn{1}{c}{$dn/dz$+Fup} & \multicolumn{1}{c}{$dn/dz$+Both}  \\
\hline
\multicolumn{6}{l}{\emph{SPT SZE Survey}}\\
\hspace{5pt}$\Delta$\om				& 0.0084 & 0.0190 & 0.0237 & 0.0115 & 0.0181 & {\bf0.0108} (0.0096)\\
\hspace{5pt}$\Delta$$\sigma_8$ 	         & 0.0105 & 0.0236 & 0.0441 & 0.0196 & 0.0218 & {\bf0.0125} (0.0111)\\
\hspace{5pt}$\Delta$$w$ 	         		& 0.0371 & 0.0663 & 0.1745 & 0.1543 & 0.0602 & {\bf0.0376} (0.0335)\\
\hline
\multicolumn{6}{l}{\emph{$Planck$ SZE Survey}}\\
\hspace{5pt}$\Delta$\om				& 0.0078 & 0.0575 & 0.0820 & 0.0188 & 0.0344 & {\bf0.0062} (0.0059)\\
\hspace{5pt}$\Delta$$\sigma_8$ 	         & 0.0123 & 0.1097 & 0.1747 & 0.0393 & 0.0629 & {\bf0.0098} (0.0094)\\
\hspace{5pt}$\Delta$$w$ 	        		 & 0.0397 & 0.1977 & 0.3930 & 0.1049 & 0.1149 & {\bf0.0404} (0.0390)\\
\hline
\multicolumn{6}{l}{\emph{DUET X-Ray Survey}}\\
\hspace{5pt}$\Delta$\om				& 0.0099 & 0.0290 & 0.0372 & 0.0114 & 0.0142 & {\bf0.0096} (0.0083)\\
\hspace{5pt}$\Delta$$\sigma_8$ 	         & 0.0131 & 0.0503 & 0.0846 & 0.0127 & 0.0182 & {\bf0.0114} (0.0100)\\
\hspace{5pt}$\Delta$$w$ 	       		 & 0.0505 & 0.0806 & 0.3540 & 0.2065 & 0.0765 & {\bf0.0625} (0.0527)\\
\hline
\end{tabular}\hfil}
\end{table*}

\section{Cosmological Parameter Constraints}
\label{sec:results}
Our main results are summarized in Table 1 and highlighted in the following three figures, one for each survey. The table contains the 1-$\sigma$ parameter uncertainty for each survey for \om, $\sigma_8$
and $w$ and addresses the main focus of this paper, which is to examine the degree to which one can tighten cosmological from the redshift distribution $dn/dz$ by adding the cluster power spectrum $\bar{P}_{cl}$ and detailed followup of 100 clusters.   With the exception of \citet{majumdar03} and \citet{hu03b}, previous studies of cluster surveys have implicitly assumed full knowledge of cluster structure evolution and only addressed the 
the sensitivity of cluster surveys to cosmological parameters. Constraints from such considerations
are grouped under the heading `Only Cosmology'  in Table 1. However, if one does not assume
perfect knowledge of cluster structure and lets the survey solve for both cosmology and cluster 
scaling parameters, then the cosmological constraints are weakened. This is illustrated in the next column labeled `Std Evol.'. This approach, however, still assumes that there is no unexpected evolution in  scaling relation with redshift and ignores any uncertainty due to it. A cautious approach would be 
to try to solve for any evolution from the survey itself (see Eqn \ref{eq:lx-m} and \ref{eq:fsz-m}). The
estimated uncertainties for this self-calibrating case are shown in column labeled `Self-Cal'. The next column shows constraints when one combines information from both $dn/dz$ and $\bar{P}_{cl}$, both calculated for
the self-calibration case. The second last column shows the constraints if one does a mass 
follow-up of 100 clusters from the survey sample and the final column shows the cosmological
constraints when one uses information from $dn/dz$, $\bar{P}_{cl}$ and also follow-up. The constraints are using cluster surveys only. However, in the last column, within parentheses, we show an estimate of how the  constraints change if one includes strong priors of \om, $\sigma_8$ and $w$.

From the table, one is able to draw two general conclusions: 
\begin{itemize}

\item Assuming complete knowledge of cluster 
structure can give overly optimistic constraints on cosmological parameters. A complete lack of
knowledge of cluster structure can drastically degrade our ability to constrain such parameters 
(especially $w$) from future cluster surveys.  This is illustrated in columns 2-4.

\item The last three columns show that one can greatly improve the constraints if one uses information from power spectrum studies or mass follow-up of 100 clusters.   Using all available information from $dn/dz$, $\bar{P}_{cl}$ and follow-up can deliver uncertainties comparable to the case of using the redshift distribution alone and assuming perfect knowledge of cluster structure
and evolution. This, indeed, is very encouraging.

\end{itemize}

\subsection{Self-calibration and degradation of constraints}
For the case with perfect knowledge of cluster structure, all the three surveys compare favourably,
yielding 1-$\sigma$ uncertainty on $w$ of 0.0371, 0.0397 and 0.0505 for the SPT, $Planck$ and DUET
surveys respectively. The marginally better uncertainty on $w$ from SPT over $Planck$ is due to the higher sensitivity of the SPT survey, which delivers higher redshift clusters that are 
more sensitive to change in
 $w$ (for example, see Figure 4 in MM03). $Planck$, which essentially detects only high mass clusters over the entire extragalactic sky, constrains $\Omega_M$ most effectively. Note, that when we have no uncertainties entering from any imprecise knowledge of cluster structure, SPT having the
maximum number of clusters and the deepest redshift range best constrains $\sigma_8$. However,
with the addition of extra information from the cluster power spectrum and mass follow-up,  the shallower but largest solid angle $Planck$ survey provides the  the best constraints on $\sigma_8$.
The  joint 1-$\sigma$ constraints for this case on $w$ and \om are shown in Figure  \ref{fig:w-om} and on $\sigma_8$ and \om in Figure \ref{fig:sig-om} using dotted lines.

When one solves for both the normalization and the slope of cluster scaling relations,
along with cosmology, from cluster surveys,  the uncertainties degrade by factors of 
upto 2-5 depending on the survey and the parameter concerned. The fact that one still obtains uncertainties on $w$
of 0.0663 and 0.0806, for SPT and DUET respectively, shows that it is possible to constrain both cosmology and cluster physics
reasonable well from future cluster surveys \citep[see also][]{levine02}.  The weakening of parameter constraints is most significant for $Planck$ 
for which  the peak of $dN/dz$ falls at $z \sim 0.14$ which is much below the modal redshift in $SPT$ or
$DUET$.
This is due to fact that lower sensitivity surveys that are picking out only the few, most massive clusters can less effectively constrain the cluster mass-observable relation normalization and slope (i.e its more costly to solve for cluster structure) and these larger uncertainties lead to weakened cosmological parameter constraints (for example, $Planck$ can constrain the slope of the mass-observable relation to  $\sim 27\%$ compared to $\sim 17\%$ and $9\%$ from DUET and SPT). 

The crux of the problem arises when one takes into account the possibility of non-standard
evolution. When one requires that the survey also solve for the evolution of the mass--observable relation, the constraints on $w$ weaken {\it further} by a factor of $\sim 2-2.5$ 
for the two SZE surveys (i.e $\sigma_w$ $\sim$  0.1745 for SPT and $\sim$ 0.3930 for $Planck$) and a factor $\sim 4.5$ 
  for DUET. Because this uncertain evolution affects the mass estimates for high redshift clusters, it affects the constraint from DUET  more than those from $Planck$, even though both have similar total number of clusters.  Constraints on \om and $\sigma_8$ also weaken
 significantly.  The reason for this weakening of constraints is the fact that constraints on the evolution parameters  (\gamx/\gams) are very weak and this large uncertainty in the evolution of the mass-observable relation makes it harder to know the masses of high redshift clusters;  this drives the weakened sensitivity to other cosmological parameters (for additional discussion, see
MM03).  The interplay between redshift depth and total number of clusters is complex, though, for SPT, though deeper than DUET, has slightly less degradation in errors due to it having $\sim 30\%$ more clusters than DUET.  In Figures \ref{fig:w-om} and \ref{fig:sig-om}, we have shown  constraints on $w-\Omega_M$ and  $\sigma_8-\Omega_M$ for `Self-Cal' using long--dashed lines.

\myputfigure{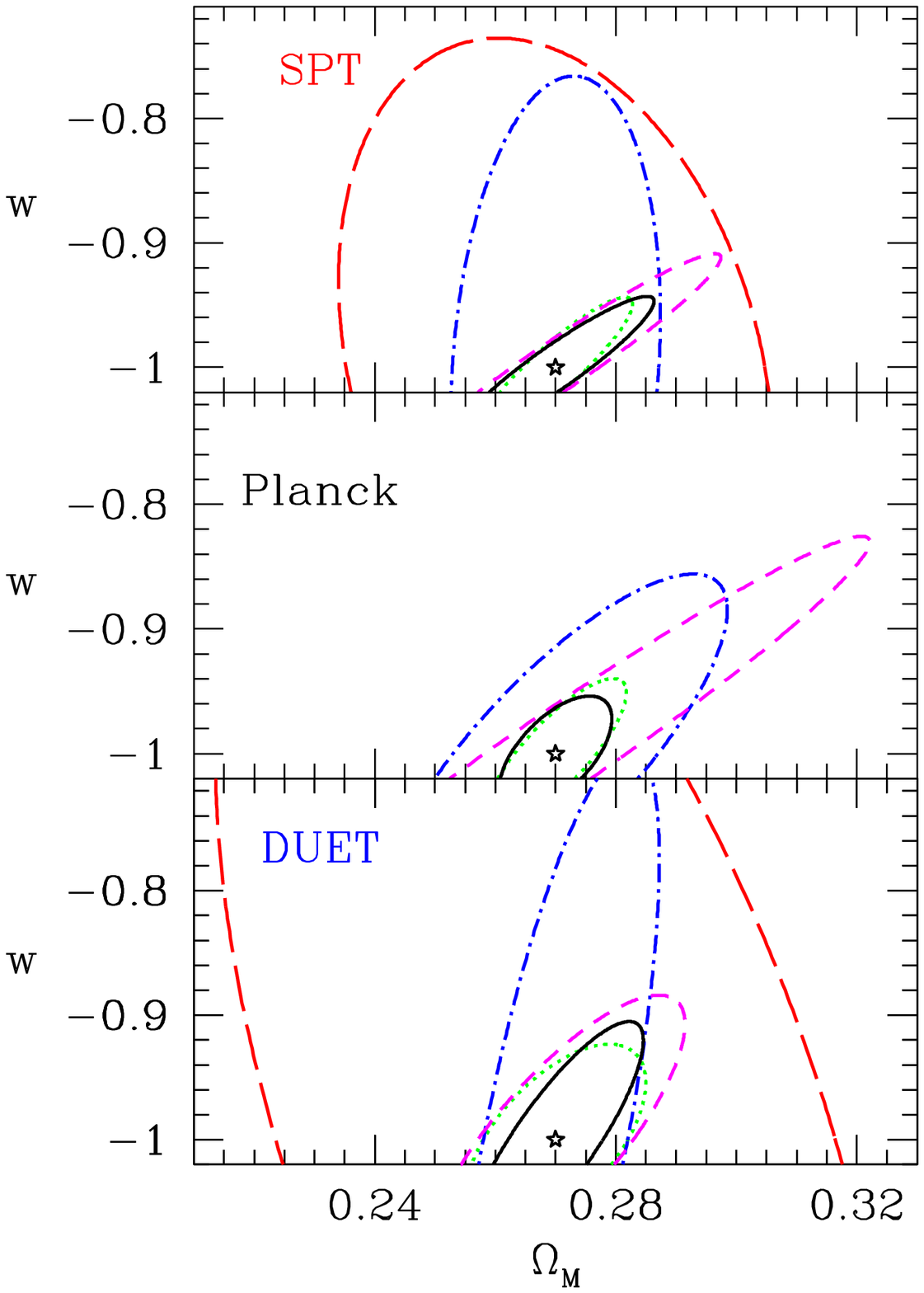}{1.9}{0.7}{-15}{-05}
\figcaption{Constraints on $w$ and $\Omega_M$  for an SZE SPT survey (top), SZE Planck survey (middle) and 
an X-ray DUET survey (bottom).   Contours denote joint 
1$-\sigma$ constraints in five scenarios: constraints from $dN/dz$ for `Only Cosmology' case (i--dotted) ;
constraints from $dN/dz$ for `Non-Std Evol' case (ii--long--dashed);  
constraints from $dN/dz + \bar{P}_{cl}$, (iii--dot--dashed); $dN/dz$ + 100 cluster follow-up (iv--short--dashed) and the 
finally constraints from
$dN/dz + \bar{P}_{cl}$ + follow-up (v--solid).  
Note, that for the Planck survey, for the `Non-Std Evol' scenario (i.e long--dashed), the constraint ellipse is too large
and lies outside the range for the two axes.  A flat universe is assumed for all the cases. The different terms are described in the
text.
\label{fig:w-om}\vskip5pt}

\subsection{Including the Cluster Power Spectrum} 
The addition of constraints from $\bar{P}_{cl}$ from the survey sample to that from $dn/dz$ provides us with a 
rather neat and effective option to dramatically improve the cosmological constraints. The fact that 
parameter degeneracies on cosmological parameters from the redshift distribution and the spatial two point
correlation function $\xi(r)$ are quite different has been demonstrated for future surveys
\citep{refregier02}. Because $P(k)$ and $\xi(r)$ are Fourier transform pairs, this
also holds for $dn/dz$ and $\bar{P}_{cl}$ as well.  Including the cluster power spectrum helps in
two ways. First,  because the amplitude of the cluster power spectrum depends on the cluster bias,  a measure of the power spectrum essentially provides a handle on the masses of the tracers 
used to measure the power spectrum. This information provides an indirect measurement of of masses of the clusters for any
particular survey and mass-observable relation.  Secondly, the shape of the power spectrum reflects the transfer function for density perturbations, which is very sensitive to the epoch of matter--radiation equality; thus, the cluster power spectrum places strong constraints on cosmological parameters, mainly $\Omega_M$.  Because of the complex parameter degeneracies,  improved constraints on \om are reflected by tighter constraints on other parameters.
 
We emphasize that any survey that delivers the redshift distribution naturally gives the spatial position and the redshift of the  tracers; hence the cluster power spectrum comes without any extra observational effort.  Although serendipitous cluster surveys may be adequate for measuring the cluster redshift distribution, it will require contiguous coverage to keep survey window functions sufficiently simple to make robust estimates of the cluster power spectrum.

It is clear from Table 1 that adding the $\bar{P}_{cl}$ constraint reduces uncertainties on $w$ for the two larger area surveys by factor of $\sim 2-4$ to make the uncertainty closer to ones obtained from the redshift distribution in the `Only Cosmology'
scenario. Using the cluster power spectrum tightens the constraint on $w$ most for $Planck$, followed by DUET and SPT. This is expected 
because of the large physical scales probed by the all sky $Planck$ survey.  This is more evident when one compares DUET with $Planck$,
both of which have similar total number of clusters.  Thus, when one combines $\bar{P_{cl}}$ information to that from $dN/dz$,
one obtains 1-$\sigma$ constraint of $\sim 10\%$ on $w$ from $Planck$ which is a factor of $\sim$4 improvement on the $\sim 39\%$
 constraint on $w$ obtained in the `Self-Cal' scenario.  Notice that addition of the power spectrum also very effectively tightens the
constraint on $\Omega_M$, especially for $Planck$, which is reduced by a factor of $\sim 4.5$ from 
$\sigma_{\Omega_M} = 0.082$ to $0.0188$.  For SPT and DUET, the $\Omega_M$ constraint also improves by factor of $\sim 2-3$.

A point to note is that the amplitude of the power spectrum, which depends on the bias of the tracers and the underlying amplitude of the dark matter power spectrum-- $\sigma_8$-- is a key component which helps in tightening the constraints. The effective bias, for a particular survey, depends on the number distribution of the clusters and their masses, which-- in turn-- depends on the cosmological parameters (mainly $\sigma_8$) and the cluster structure. Hence, constraints on $\sigma_8$
and cluster structure from using the power spectrum are very degenerate.  
Finally, we should also keep in mind that $\sigma_{FKP}$ depends not only on volume but also on number density. Hence, the area and the depth of the surveys affect the uncertainty estimation in a non-trivial manner.  Constraints from adding the power spectrum to the cluster redshift distribution are shown by dot--dashed lines in Figures \ref{fig:w-om} and \ref{fig:sig-om}.  The fact that 
parameter degeneracies from the redshift distribution and $\bar{P}_{cl}$ are quite different is evident from the rotation of the elliptical joint parameter constraints when one adds $\bar{P}_{cl}$ constraints to that from $dn/dz$.

 \subsection{Including Limited Mass Follow-up}
Yet another way to reduce the cost of forcing the survey to self-calibrate (i.e. solve for cluster structure and its evolution in addition to cosmology),  is to
add information from a mass follow-up of 100 clusters between the redshifts of 0.3 and 1.2, with 10 clusters followed-up at redshift intervals of 0.1. 
At each redshift, the 10 clusters are equally divided in mass between $10^{14}h^{-1}M_\odot$ and
 $10^{15}h^{-1}M_\odot$. This is equivalent to following up less than $1\%$ of the cluster sample in each survey, and we envision this followup as mostly being available through deep, archival X-ray observations that are being accumulated by {\it XMM-Newton} and {\it Chandra} even now.  We see that  follow-up significantly reduces the uncertainties for all the three surveys. The constraints for $w$ goes down to $\sim 0.06, 0.115$ and $0.077$ for SPT, $Planck$ and DUET respectively. There are also major  improvements in $\Omega_M$ (by a factor of 1.3-2.6) and $\sigma_8$ (by a factor of 2-4.5) uncertainties. 
Essentially, follow-up helps is tightening the constraints on the unknown cluster structure and its evolution;  this translates into more accurate mass estimates for the clusters, and that
reduces the uncertainties on the cosmological parameters. 
Because SPT probes to higher 
redshift, follow-up is most valuable for this survey which contains the most high redshift clusters, whose masses are most uncertain without the follow-up information.  Short--dashed lines in Figures \ref{fig:w-om} and \ref{fig:sig-om} show joint constraints when one added information from
a follow-up of 100 clusters to that from $dN/dz$.
MM02 demonstrated that although there is some further improvement in constraints as one follows up a larger fraction of the clusters, the
 difference between following up 10\% and 100\% of the cluster population is minimal.   We remind the reader that follow-up does not explicitly depend on $\sigma_8$, $n$, $h$ and $\Omega_B$ and hence cannot constrain these parameters by itself and contains no cross-correlation of other cluster structure or cosmological parameters with these four parameters.  

\subsection{Combining $dn/dz$, $\bar{P}_{cl}$ and Follow-up }
It is interesting at this point to compare and combine parameter constraints obtained by the
two options at hand; follow-up mass measurements versus addition of $\bar{P}_{cl}$ constraints. Other than the $Planck$ survey,  having mass follow-up
helps in tightening constraints on $w$ more than adding power spectrum information. $Planck$, which surveys the maximum
solid angle, delivers power spectrum constraints that are slightly more powerful than follow-up.
However, for $\Omega_M$ and $\sigma_8$, the addition of $\bar{P}_{cl}$ constraints fares better than addition of follow-up constraints (for example, for SPT survey addition of $\bar{P}_{cl}$ improves the constraint of $\Omega_M$ by a factor of 
$\sim 2$ compared to $\sim 1.3$ when one adds the follow-up Fisher matrix).  

As a simple extension, one can add both power spectrum and follow-up constraints to that from $dN/dz$. This, then, gives us the best possible constraints on all the cosmological parameters. One ends up having $ 3.76\%, 4.04\%$ and $6.25\%$ constraints on $w$ from SPT, $Planck$ and DUET respectively. This can be compared to the overly optimistic 3.71\%, 5.05\% and 3.97\% constraints one obtains in the `Only Cosmology' case.  
For the DUET  and $Planck$ surveys,  which are more sensitive to the massive clusters compared to SPT,  the final constraints on \om and $\sigma_8$ are even better than the `Only Cosmology' case. Thus, one not only regains  any cosmological sensitivity lost due to uncertain knowledge of cluster structure and evolution, but infact does marginally better.  The best case  scenario for parameter constraints estimated by adding constraints from $dN/dz$, power spectrum  and follow-up are shown by solid lines in Figures \ref{fig:w-om} and \ref{fig:sig-om}. In Table 1, numbers in parentheses show
the result of adding CMB constraints to that from cluster surveys. In adding constraints from CMB observations to those from cluster surveys, one should add the full covariance matrix in order to get uncertainties on individual parameters as well as parameter  degeneracies. Our estimate of the uncertainties in parentheses (in Table 1) uses the CMB priors on the individual parameters and does not include information on their cross-correlations; thus, these numbers should be viewed as rough estimates.

\myputfigure{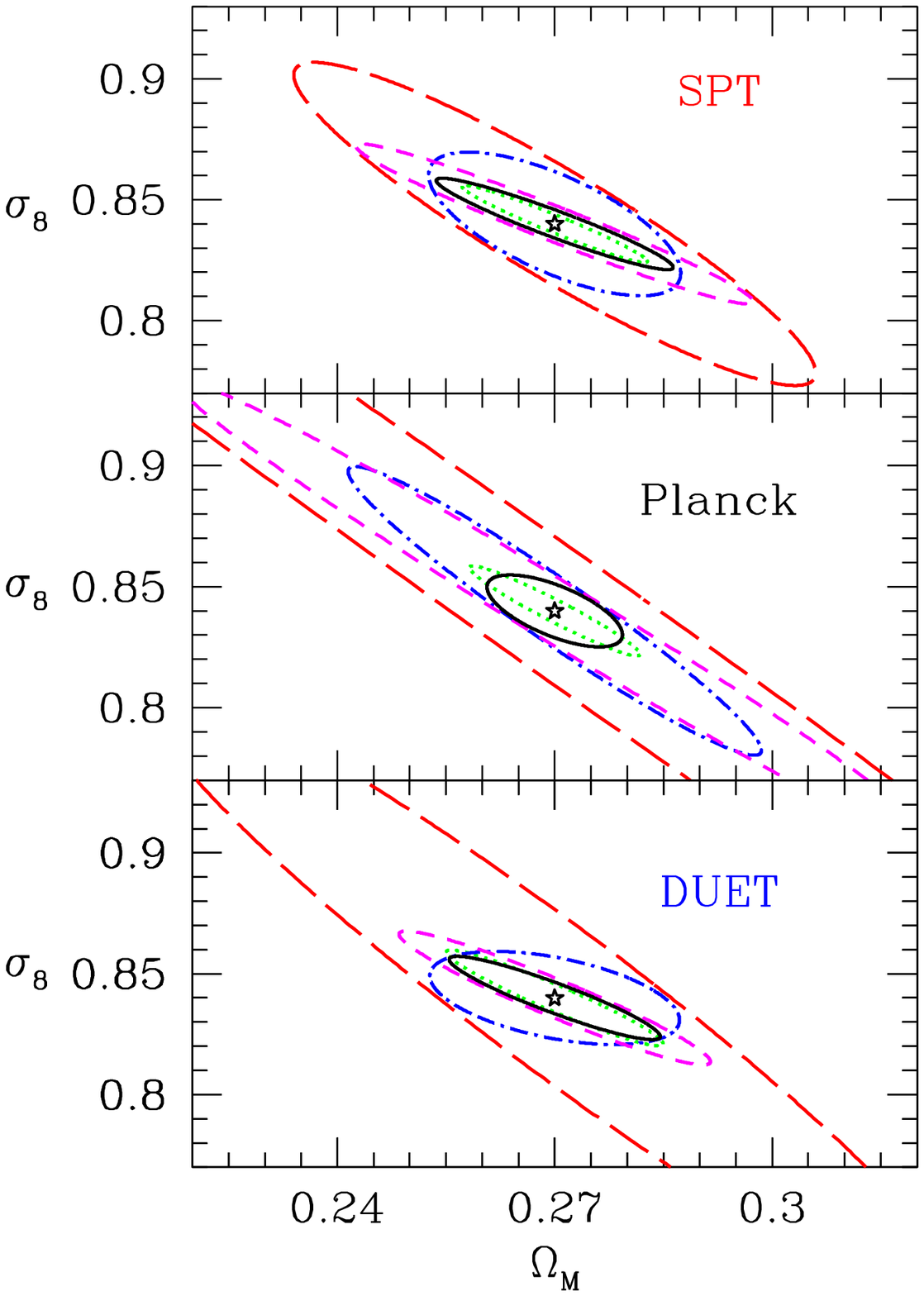}{1.9}{0.7}{-15}{-05}
\figcaption{Constraints on $\sigma_8$ and $\Omega_M$  for an SZE SPT survey (top), SZE Planck survey (middle) and 
an X-ray DUET survey (bottom).   Contours denote joint 
1$\sigma$ constraints in four scenarios: constraints from $dN/dz$ for `Only Cosmology' case (i--dotted) ;  
constraints from $dN/dz$ for `Non-Std Evol' case (ii--long--dashed);  
constraints from $dN/dz + \bar{P}_{cl}$, (iii--dot--dashed);  $dN/dz + 100$ cluster follow-up, (iv--short--dashed)and  
finally constraints from
$dN/dz + \bar{P}_{cl}$ + fo llow-up (v--solid).  A flat universe is assumed for all the cases. The different terms are described in the
text.
\label{fig:sig-om}\vskip5pt}

\subsection{Combining CMB with $\bar{P}_{cl}$ }
Cosmological parameter constraints can, of course, be derived using only the cluster power spectrum $\bar{P}_{cl}$; however,  taken by itself, the cluster power spectrum constraints are quite weak due to  large parameter degeneracies. 
Because the shape of the power spectrum depends on combinations of $\Omega_M$, $h$ and $\Omega_B$, the best
constraints are on these parameters and other parameters like $\sigma_8$ and $w$ are quite weak.
This trend has already been noted in related works using correlation functions \citep[see][]{refregier02}.
The cluster power spectrum, by itself, is unable to constrain the cluster structure and evolution parameters. Of the three surveys, $Planck$ constrains the cluster parameters a few orders of magnitude better than the other surveys  (although still extremely weakly; for example they are about a factor $\sim 5$ weaker than those from $dN/dz$ alone).
One can break these degeneracies by adding CMB anisotropy constraints and cross-correlations. 
Infact,  if one puts in WMAP CMB priors on $\Omega_M$, $\sigma_8$ and $w$, then one can get constraints on $w$ of $\sim$ 10\%, although this is not much better than the CMB prior on $w$ of 11\%. Note that in contrast, once $dn/dz$   and $\bar{P}_{cl}$ constraints  are combined, any further addition of strong external CMB priors  on $\Omega_M$, $\sigma_8$ and $w$  has little effecton  the constraints (see numbers within parentheses in Table 1). 
In fact, it's important to stress that future, self--calibrating cluster surveys by
themselves can give roughly a factor of 2.5 (e.g $4\%$ for SPT) improvement on $w$ constraints
when compared to those obtained using present CMB data \citep{spergel03}.

\myputfigure{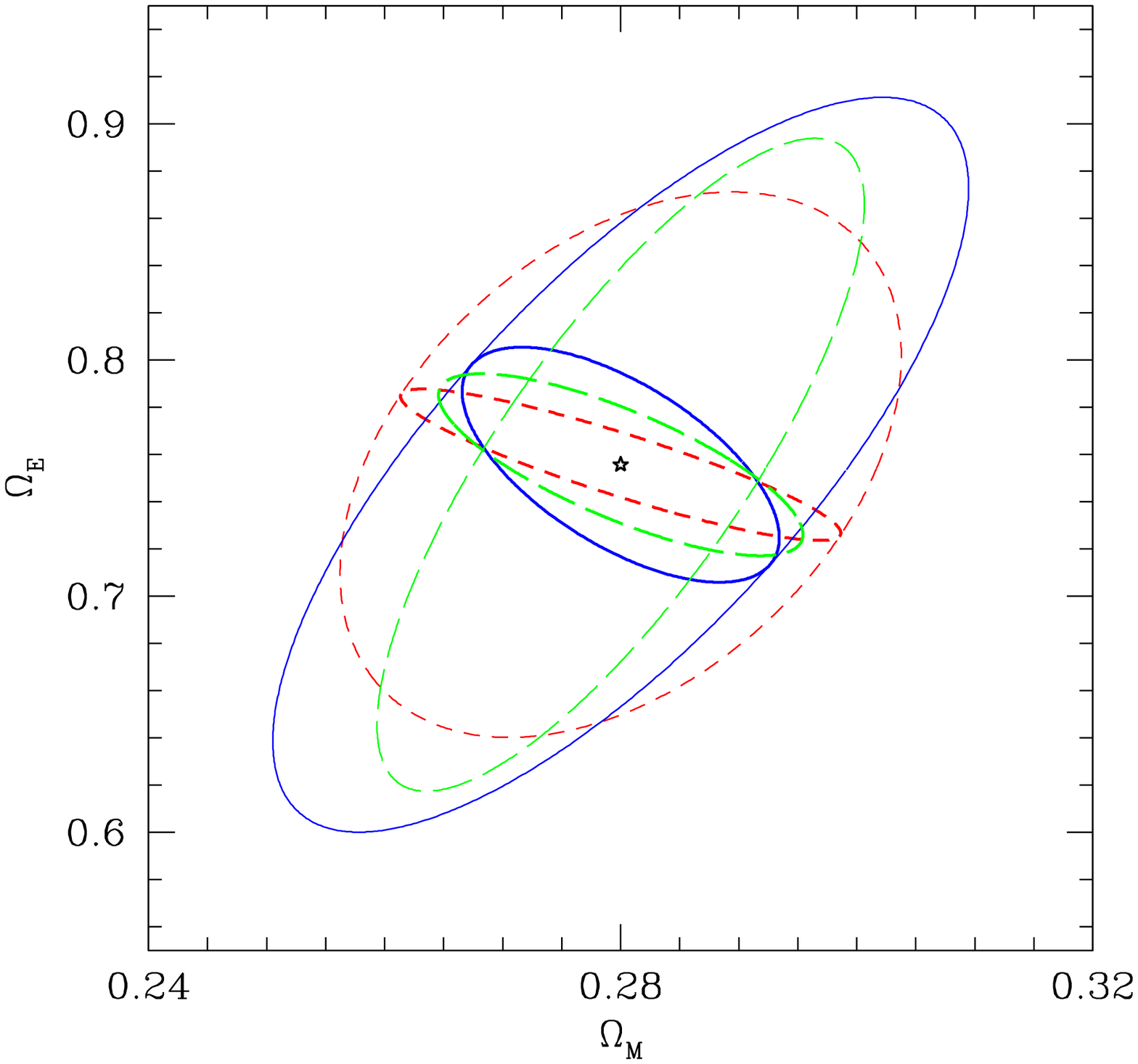}{2.85}{0.5}{-25}{-05}
\figcaption{Constraints on $\Omega_E$ and $\Omega_M$  for the three surveys; SPT (short--dashed lines),
$Planck$ (long--dashed lines) and DUET (solid lines).  The thin lines are for freely varying $w$ and the thick lines are for fixed $w=-1$. 
\label{fig:omM-omE}\vskip5pt}

\subsection{Solving for Curvature and the Equation of State }
We have also examined the effects of our flat universe prior by allowing $\Omega_{tot}$ to vary. This gives us an opportunity to examine the $\Omega_E$ constraints and the $\Omega_E$-$\Omega_M$ degeneracy (shown in Figure \ref{fig:omM-omE}). We look at both
cases of a freely varying $w$ and a fixed $w=-1$. We find the best constraint on
$\Omega_E$ to be given by the SPT survey, and these are $\sigma_{\Omega_E} = 0.0667 (0.0185)$ when $w$ is free ($w$  fixed), followed by $Planck$ and then DUET. When we forego the assumption of a flat  universe and also solve for the curvature along with the all the other parameters, we see only marginal degradation in the
$\Omega_M$ or $\sigma_8$ constraints. As an example, for $Planck$, the respective constraints are 0.0102 and 0.0211 when curvature is free. On the other hand, the $w$ constraints are much affected, and they degrade by factors greater than 3 to
$\sigma_w = 0.1353, 0.1092$ and $0.1955$ for SPT, $Planck$ and DUET, respectively.
Note that   $\Omega_E$ constraints from these surveys are better than the WMAP constraints,  and  although the $w$ constraints have weakened, they are also comparable to or better than constraints the currently available tightest constraints that come from a combination of WMAP CMB measurements together with the galaxy power spectrum \citep{spergel03} .  We note that the parameter degeneracies are potentially quite complementary to both CMB anisotropy and distance measurements, and thus one can obtain tighter constraints when one combines these different probes of dark energy. 

\section{Discussion and Conclusions}
\label{sec:conclusions}
Any attempt to precisely measure cosmological parameters, especially $w$, from future cluster surveys
will require an understanding of the mass-observable relations for clusters and the evolution of these relations.  This is true whether clusters are found using X-rays, the SZE distortion, light from the galaxies or the weak lensing shear. Here we examine the effect of current uncertainties about cluster structural evolution for three future surveys and require that information from the surveys be used to calibration the structure and evolution of clusters in addition to solving for cosmology.  We show that  the process of self-calibration weakens cosmological constraints by a factors of $4-5$, even for the deepest survey,  in comparison to constraints derived by assuming perfect knowledge  of cluster structure and evolution. 

We have demonstrated that adding information from the redshift averaged cluster power spectrum to that from cluster counts would help in reducing the constraints up to factor $\sim$ 4.  Measurement of cluster power spectrum helps us get a handle on the the masses of the tracer particles through the mass dependence of the effective bias $b_{eff}$. Moreover, the cluster power spectrum delivers an independent measure of the parameters important to the shape of the transfer function-- primarily $\Omega_M$, $h$ and $\Omega_B$, and the parameter degeneracies are also different to those from $dN/dz$. Estimation of power spectrum from a cluster survey designed to measure the redshift distribution of clusters would not require any additional observational effort and hence comes rather cheaply. However, to get the best power spectrum constraints, one needs to survey large, contiguous
regions of the sky.

We have also examined the utility of detailed mass followup observations of 100 clusters, distributed in a redshift range of  $0.3\le z\le1.2$.  This followup information is being enabled by deep {\it Chandra} and {\it XMM-Newton} observations  today, and this information will help determine the parameters of cluster mass--observable relations and their evolution.

The best constrains on the nature of dark energy and other cosmological and cluster parameters are obtained when the cluster redshift distribution is combined with the cluster power spectrum and the mass follow-up.  One can achieve $\sim$ 4\% constraint on $w$ for the two SZE surveys. This is a factor of 2.5 better than the best currently available constraints \citep{spergel03}. We emphasize that this constraint is from a self-calibrating cluster survey, where the cluster mass-observable relation and its evolution are pulled directly from the survey and followup.  Recently, \citet{hu03b} has shown that by including the shape of the mass function within survey redshift bins one can essentially solve for the cluster mass-observable relation as a function of redshift-- without including any information from the cluster power spectrum or follow-up mass measurements.  Clearly, the best approach will include all the available information-- the mass function, redshift distribution, cluster power spectrum and detailed mass measurements.

In constructing $F^p_{ij}$, we have subdivided each cluster sample into three redshift bins such
that each bin has equal  number of clusters. It is important to have thick redshift bins so as to avoid
cross-correlation between adjacent bins. The final constraint would depend to a small degree on the binning of the cluster sample.  For example, for SPT, if we subdivide the clusters in redshift bins $z=0-0.3, 0.3-0.7,0.7-3.0$, then $\sigma_w$ from $dN/dz + \bar{P_{cl}}$ is $0.1577$ and very close to $0.1543$ which was obtained when the clusters were divided to have equal number in each bin. Incidentally, if one has a single bin from $z=0-3.0$ rather than three redshift bins,  one gets $\sigma_w = 0.1568$. However, depending on the surveys, optimal redshift binning and the use of the full covariance matrix are important issues to consider in subsequent works. 

To conclude, in the present paper we have demonstrated the self--calibrating nature of cluster surveys and shown that the cluster survey is a powerful technique for studying the nature of the dark energy-- as well as any other characteristic of the universe that affects the expansion history, growth of density perturbations or the nature of the transfer function.  Constraints from cluster surveys tend to exhibit parameter degeneracies that are quite different from other techniques such as CMB anisotropy studies or distance measurements.  By pursuing all these complementary means of studying the nature of dark energy and dark matter, we will produce independent constraints that will serve as cross checks of the different systematics underlying each technique.  In addition, because the cluster survey is generating cosmological constraints from different physical properties of the universe than other techniques (i.e. surveys probe the growth of density perturbations whereas distance measurements only probe the expansion history), the pursuit of these different techniques makes it possible to precisely test the current framework for understanding our evolving universe.  We look forward to the results of these precise, new tests with great anticipation.

\acknowledgments
We wish to thank Zoltan Haiman for many conversations that have positively
impacted this study.  SM wishes to thank Y-T Lin for technical help
and Ben Wandelt for many useful discussions.  JM wishes to thank the
DUO collaboration for constructive comments during the later
stages of this work.  JM acknowledges financial support from NASA
Long Term Space Astrophysics award NAG 5-11415 and an NSF Office of   
Polar Programs award OPP-0130612.

\bibliographystyle{../../Bib/Astronat/apj}
\bibliography{../../Bib/cosmology}
\end{document}